\documentclass[pof,aps,superscriptaddress]{revtex4-1}

\pdfoutput=1
\usepackage{amsmath}    
\usepackage{graphicx}   
\usepackage{verbatim}   
\usepackage{color}      
\usepackage{subfigure}  
\usepackage{hyperref}   
\begin{document}

\title{Smartphone Schlieren}
\author{Victor A. Miller}
\affiliation{Moonlight Labs, Palo Alto, CA 94301}
\author{Keith T. Loebner}
\affiliation{Moonlight Labs, Palo Alto, CA 94301}
\affiliation{Stanford University, Stanford, CA 94305}
\date{September 13, 2016}

\begin{abstract}
We present a schlieren system comprised of 3D printed optical mounts, a sub-\$10 mirror, and a smartphone camera.  The system is intended to make schlieren imaging accessible to K-through-12 students, educators, as well as hobbyists.  In the manuscript, we detail the design of the system, provide source files for continued iteration, and show some example schlieren images and videos of a butane lighter, a jet of compressed air, and an electric stove.    
\end{abstract}

\maketitle

\section{Introduction}

Schlieren imaging is a widely used scientific imaging technique. By generating contrast in regions with varying index of refraction, schlieren imaging provides researchers the ability to visualize a variety of phenomena that are otherwise invisible or  difficult to see, like shock waves, thermal plumes, and optical imperfections. Typically, a schlieren set up consists of a light source, some type of an optical cutoff, and a camera to capture images. The optical cutoff (e.g., a knife edge) blocks or passes rays that are refracted by a change in index of refraction, thereby generating contrast. Gary Settles, the contemporary king of schlieren imaging, has written a wonderful text detailing the underlying principles and highlighting many different flavors and configurations of schlieren (and shadowgraph) imaging \cite{settles2012schlieren}.  

Schlieren images are often very pleasing to see, in part because they provide insight into otherwise invisible phenomena.  Examples of this include imaging the thermal plume of a candle (first explained as shadowgraph by Hooke, and subsequently detailed by Toepler \cite{toepler1906beobachtungen}) and the wake of a supersonic aircraft \cite{weinstein1994optical}.   This ability to elucidate otherwise invisible phenomena makes schlieren images powerful, especially when the subject is interesting or relatable \cite{nytimes_match}.

We observed the educational power of schlieren firsthand while teaching graduate students about optical diagnostics. From 2011 until 2014, we taught a laboratory unit on schlieren; the unit consisted of one three hour block of experiments in the lab and a week to process the acquired image data; the unit culminated in a 10 minute presentation on the findings. During the lab, we used a high speed camera and schlieren imaging to study an oblique shockwave generated over a wedge in an expansion tube; students were asked to process the images and algorithmically extract the shock angle, from which they inferred the Mach number of the flow. Generally, students were receptive to this lab unit, and they seemed to prefer this lab unit compared to the complementary units focused on other gas sensing and imaging techniques.

In 2015, we modified the shockwave imaging lab unit to instead image the laser-breakdown-induced ignition of a bubble filled with a mixture of methane and air. We asked students to process the images to extract the flame speed, which can be used to infer fuel-to-air ratio. The videos we acquired were quite dramatic \cite{nytimes_bubble}, and students were far more engaged with this new unit compared to the shockwave imaging unit.  We received unsolicited, glowing feedback from the students about how much they enjoyed and learned from the unit.

From this anecdotal experience, we were curious as to whether schlieren imaging can be made more accessible and turned into a teaching aid, specifically for K-12 education. Such a system would need to be easy to use, inexpensive, and accessible. An easy-to-use schlieren imaging system could be useful in a number of contexts: the system could be used as an aid to teach some concepts in fluid mechanics, like buoyancy and turbulence; the system could also be used as a platform to teach image processing, as schlieren images can often be improved with some post processing like unsharp masking; and concepts in optics, rays, and waves could be taught using the schlieren rig itself, by giving students the opportunity to understand how the acquired images change as optical elements are moved.  Lastly, and perhaps most importantly, we hope that by giving students the ability to acquire their own schlieren images, we can instill interest, curiosity, and excitement about science and engineering.

Improving the accessibility of schlieren imaging is not a new idea \cite{settles2012schlieren,kleine2013schlieren}.  Furthermore, using smartphones for scientific imaging is also not a new concept; recently, work was published detailing the use of a smartphone camera for particle image velocimetry (PIV) \cite{Cierpka2016}.  There are a variety of internet- and kit-based learning aids that promote and facilitate at-home, build-it-yourself pedagogy; for example, a recently available kit developed through a Pixar and Kahn Academy partnership (details can be found at \url{https://www.khanacademy.org/partner-content/pixar}) aims to teach kids a slew of concepts in computer graphics, modeling, and imaging.

\begin{figure}
	\includegraphics[height=75mm]{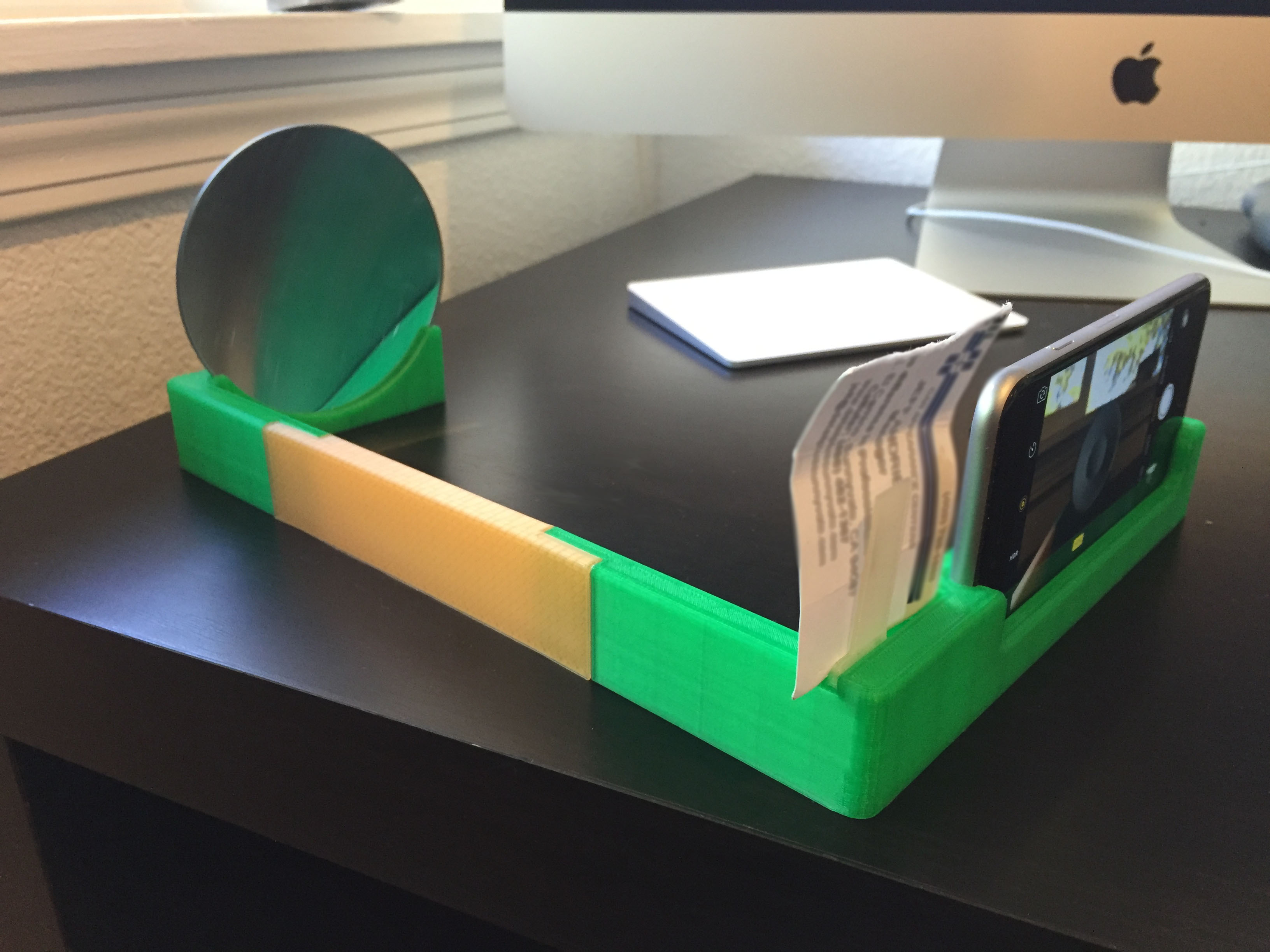}
	\includegraphics[height=75mm]{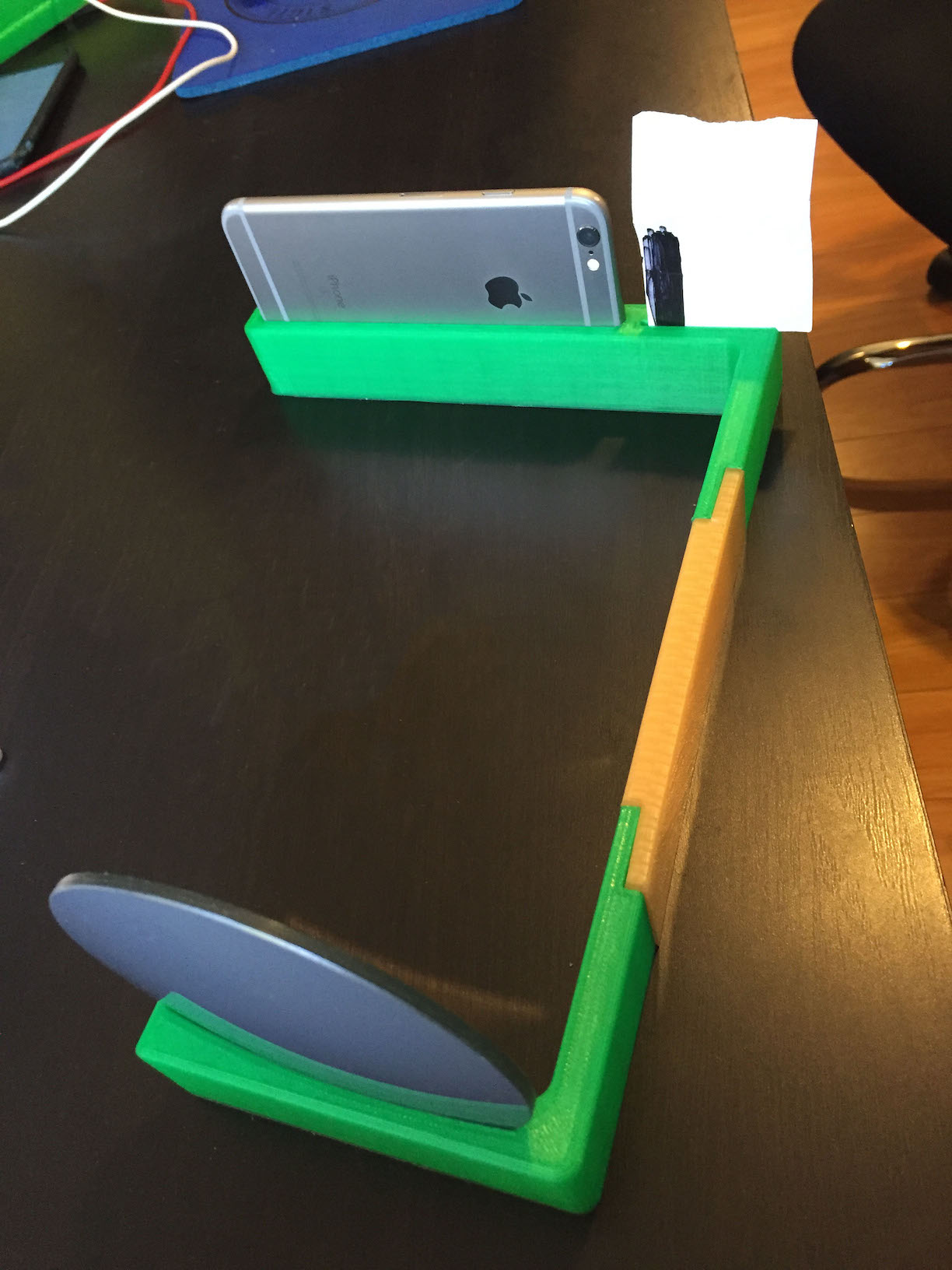}
	\caption{\label{fig:system_photo}The assembled 3D printed components of OpenSchlieren}
\end{figure}

So, in order to make schlieren more accessible for K-12 education, we've developed a cheap, 3D printed schlieren system that uses a single, inexpensive mirror, and a smart phone for the camera, which is shown in Figure \ref{fig:system_photo}.  The 3D printed mount reduces the burden of alignment on the user, and material cost, excluding the phone, is under \$10 (USD). Because of the low cost, we have considered naming the system, endearingly, as cheapOschlieren.  Other names under consideration more seriously are Flowstagram and OpenSchlieren; henceforth, for now, we will refer to the system as OpenSchlieren.  In the following sections, we detail the design of the system, present some specifics on fabricating the system, describe set up and use, and finally show some results.  All solid models (.stp) and 3D printing files (.stl) are available upon request.     

\section{Design Process}
Throughout our design process, we worked in the context of K-12 educators and students, aiming to maximize usability and ease of setup while minimizing cost.  We considered both the overall optical configuration (e.g., a z-type system \cite{settles2012schlieren}) as well as the detailed design, fabrication, and components of the system.

Through this process, we recognized that one of the most challenging aspects of setting up a schlieren imaging system is alignment, which obviously impacts usability and ease of setup.  The difficultly of system alignment makes schlieren unapproachable even to some seasoned researchers.  So, for OpenSchlieren, we aimed to reduce alignment burden by minimizing the degrees of freedom within the system by designing optic mounts that locate all of the optical elements in fixed positions.  3D printing is particularly useful for producing these optical mounts, especially now that consumer-grade fused deposition molding (i.e., FDM 3D printing) is reasonably accessible to students and educators. However, FDM printers can produce parts with large tolerances resulting from part warpage and performance differences between printers. Furthermore, the small print volume typical of consumer grade FDM printers  (e.g., 8 $\times$ 8 $\times$ 8 inches) limits the size of the schlieren system, which also has implications for our optics selection.  So in summary, 3D printed schlieren fixtures provide a cheap and easy way to produce fixtures for aligning the system, but the design of these fixtures must also afford the user the ability to make adjustments to compensate for large production tolerances, all while still being easy to align.

Optic selection plays a role in usability and cost.  Schlieren optics, specifically the collimating mirrors or lenses, can be expensive.  As optics increase in size, the larger and more usable the field of view becomes, but large aperture optics are typically more expensive than similar, smaller aperture elements.  To compromise in maximizing usability while minimizing price, we chose to use the largest aperture, lowest quality and therefore least expensive optics we could source; the mirror we ultimately used is a 100mm diameter, 150mm focal length mirror which costs about \$3.50 (part number MCC106 from \url{http://www.novatech-usa.com}).

These inexpensive optics typically suffer from spherical aberration, which does not affect the resulting images beyond the point of interpretation.  At risk of being pollyannaish, the aberration may even be used as a teaching point, providing students a clear target to eliminate while post processing their schlieren images.

\begin{figure}
	\includegraphics[width=150mm]{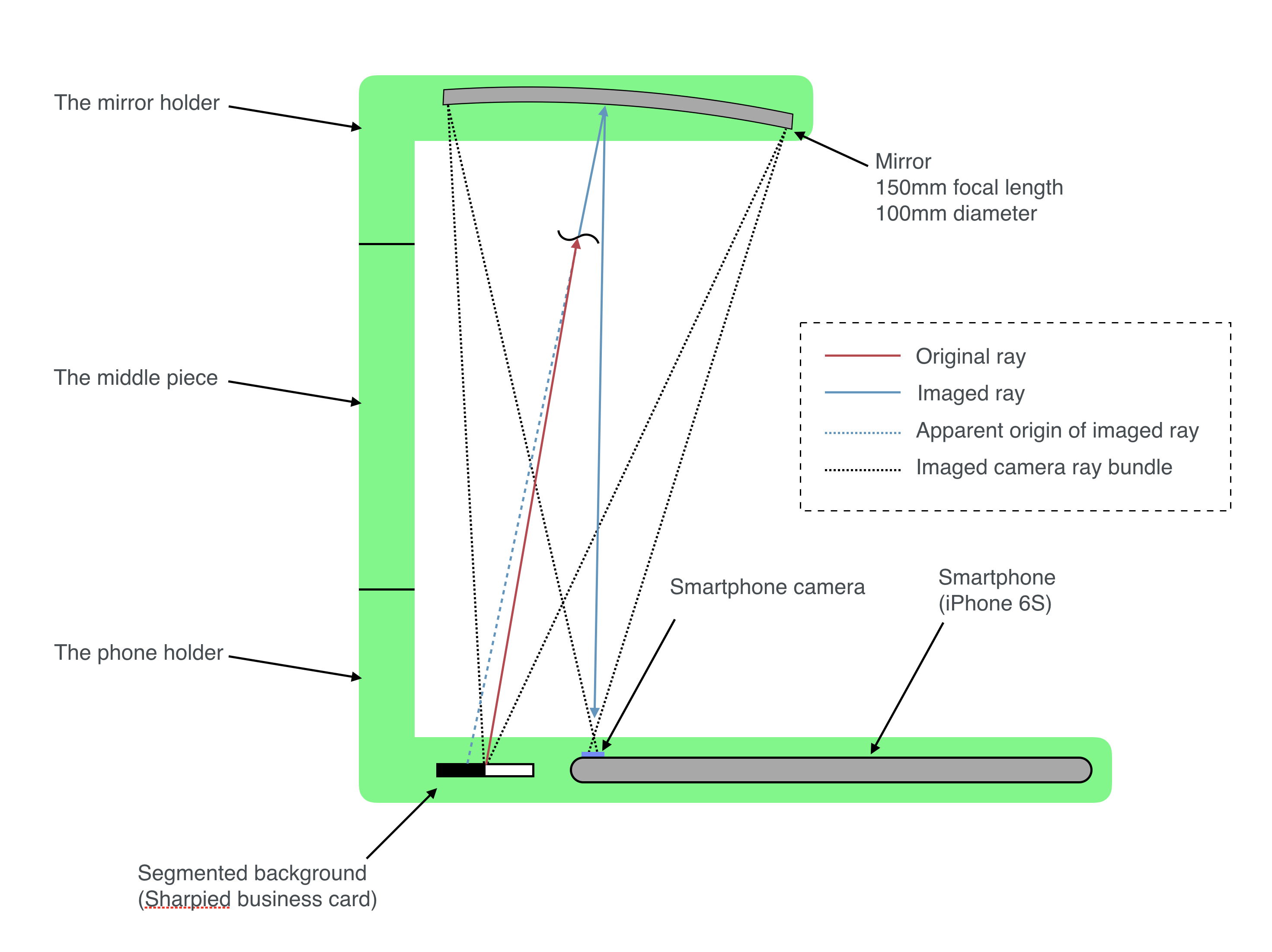}
	\caption{\label{fig:schematic} Schematic of OpenSchlieren.}
\end{figure}

From here, we surveyed a number of schlieren configurations, and settled on a single mirror, cutoff-source-image configuration, shown in Figure \ref{fig:schematic}.  This configuration consists of a single concave mirror, and a segmented background source image that acts as the cutoff; a smart phone camera captures the resulting images.  The segmented background source is a business card with a vertical black stripe drawn with a marker; the direction of this band is perpendicular to the gradients visualized by the system (i.e., a vertical band visualizes horizontal gradients).  Contrast is generated when a ray is steered so that its apparent origin falls onto the light or dark band of the source image.    

We did consider and test a conventional Z-type system with two mirrors because of its popularity in the aerospace research community. Such a system could be designed to use two separate phones, one for the LED illumination source (i.e., the smartphone camera flash) and one for the camera; a z-type system could also be designed that uses both the LED light source and camera from a single smartphone. Ultimately, we decided to go with a single mirror system because devising a 3D printed system to easily align the two mirrors in a z-type system was deemed too challenging for now.

The first single-mirror configuration we tested used the smartphone camera flash LED as the illumination source; the mirror was aligned such that the reflected beam fell on the smartphone camera; the distance from the phone to the mirror was such that the refocused beam was in front of the camera, and a knife edge could be positioned at the focus of the beam just in front of the smart phone camera lens. Aligning this system was challenging for primarily two reasons: first, the small aperture on the smartphone camera increased the difficulty of designing optical mounts that were easy to align and robust to vibrations; and secondly, using an LED for illumination and focusing that beam onto the lens saturated the camera (this was also a problem in the z-type configuration). Many smartphone cameras restrict explicit control of aperture and exposure times one could adjust to control the exposure and eliminate the saturation. Even if these aperture and exposure controls are accessible, by stopping down and decreasing exposure time in order to prevent saturation, one loses all of the ambient light and objects in the frame, which can be disorienting and, subjectively, renders the schlieren a little less approachable.

The single mirror configuration is a double pass system, so depending on the location of the refractive gradient relative to the mirror, the contrast in the image can appear twice. This results from rays being refracted twice on their way from the cutoff-source to the camera; the further away from the mirror the density refractive gradient occurs, the bigger the offset between the two density gradients. Therefore, in results presented in the next section, subjects are positioned as close to the mirror as possible to minimize the double imaging.  

\section{Design and Results}

\begin{figure}
	\includegraphics[width=150mm]{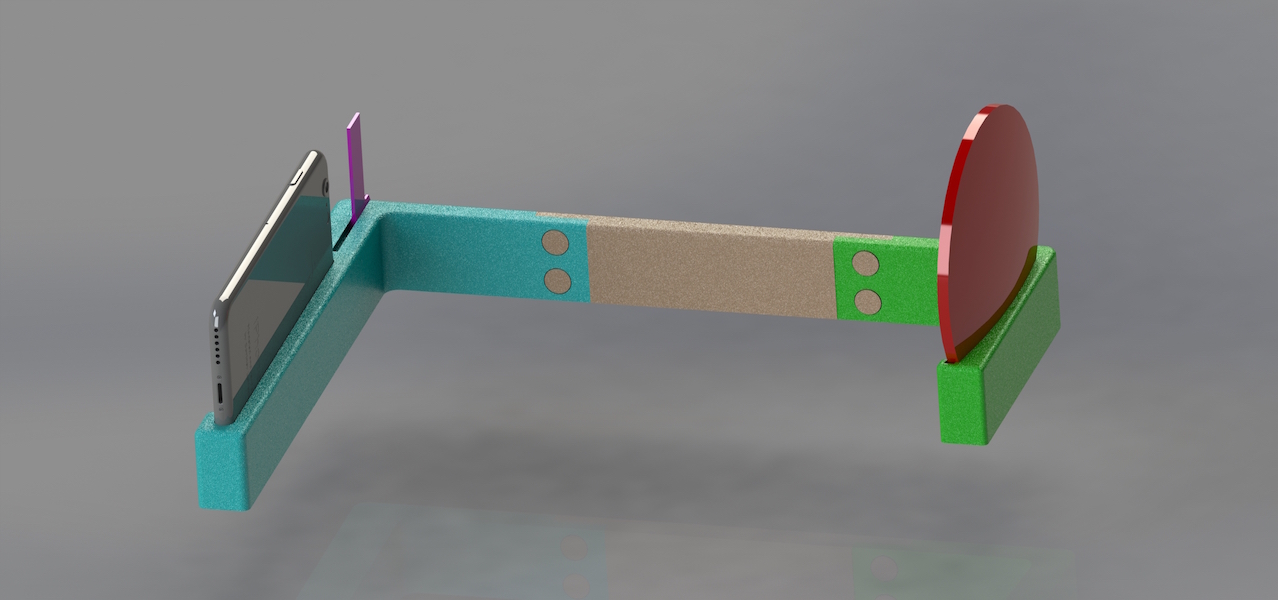}
	\includegraphics[width=150mm]{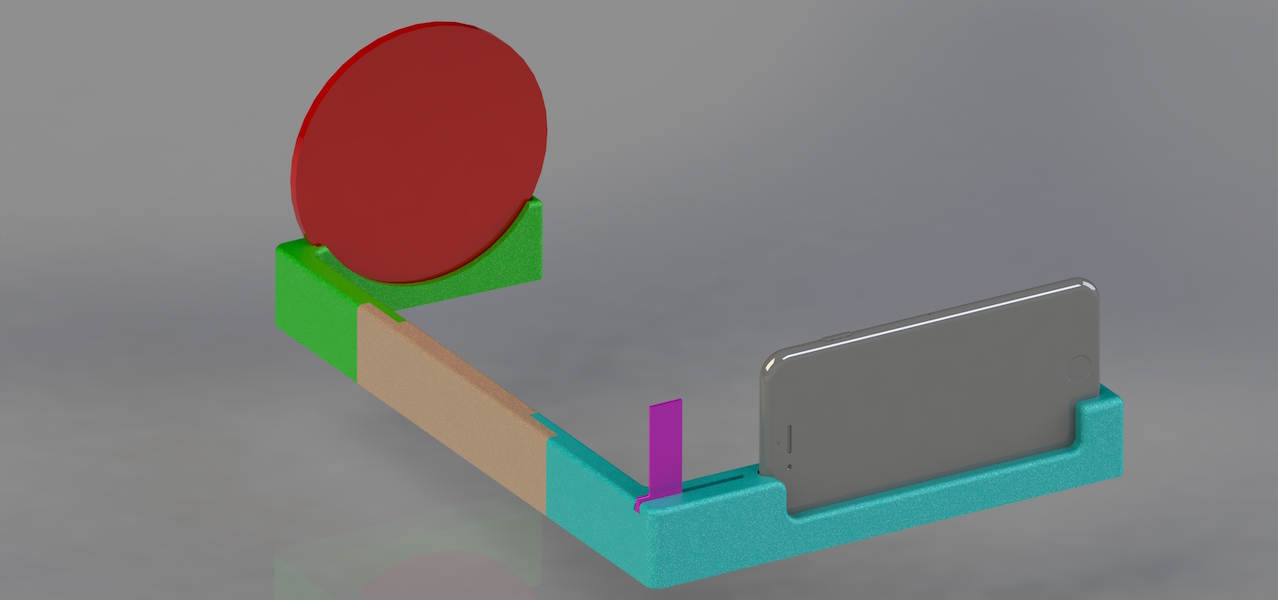}
	\caption{\label{fig:cad} 3D rendering of OpenSchlieren}
\end{figure}

The final design for this version of OpenSchlieren is shown in Figure \ref{fig:cad}.  The system consists of four separate pieces: three primary members and one piece that props up a business card as the segmented background.  Of the three primary members, one member holds the mirror (i.e., `the mirror holder', green in Figure \ref{fig:cad}), one member holds the phone (i.e., `the phone holder', blue in Figure \ref{fig:cad}), and the last member (`the middle piece', orange in Figure \ref{fig:cad}) mounts to the former two members to locate the mirror at the correct distance from the phone camera.  These three members mount together using pegs incorporated into the members; this particular aspect of the design, that is, the design for mating the different pieces together, is ripe for improvement.  The fourth piece, dubbed `the business card holder' (magenta in Figure \ref{fig:cad}), slides into the slot in 'the phone holder'.    

All four pieces were printed simultaneously on a consumer grade FDM printer.  It took about 3 hours to print all of the pieces on a medium quality setting with 0.2 mm thick layers of PLA material.  The phone holder member holds an iPhone 6 or 6S.  Others can modify this piece to interface with different phones.  The key design aspect for the phone location is that the camera must be located roughly in the same spot as the iPhone 6 or 6S is in the current design.  Roughly, the camera must be within about $\pm$ 1cm horizontally and vertically of the iPhone 6/6S cameras, and within about $\pm$ 3mm along the optical axis.  The smartphone camera must be located such that it images the focus of the mirror, where the cutoff-source-image is located (See Figure \ref{fig:schematic}).

\begin{figure}
	\includegraphics[width=50mm]{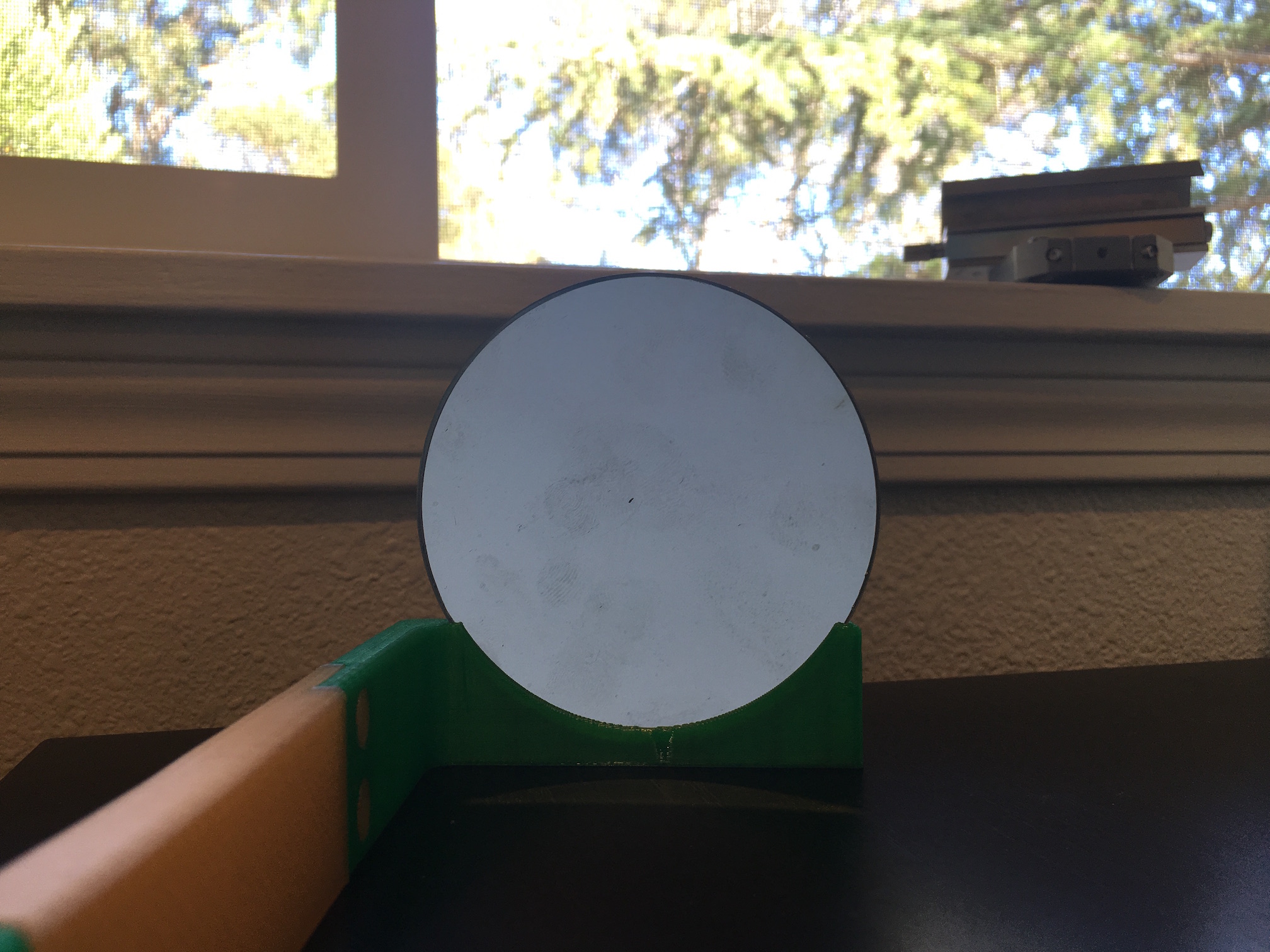}
	\includegraphics[width=50mm]{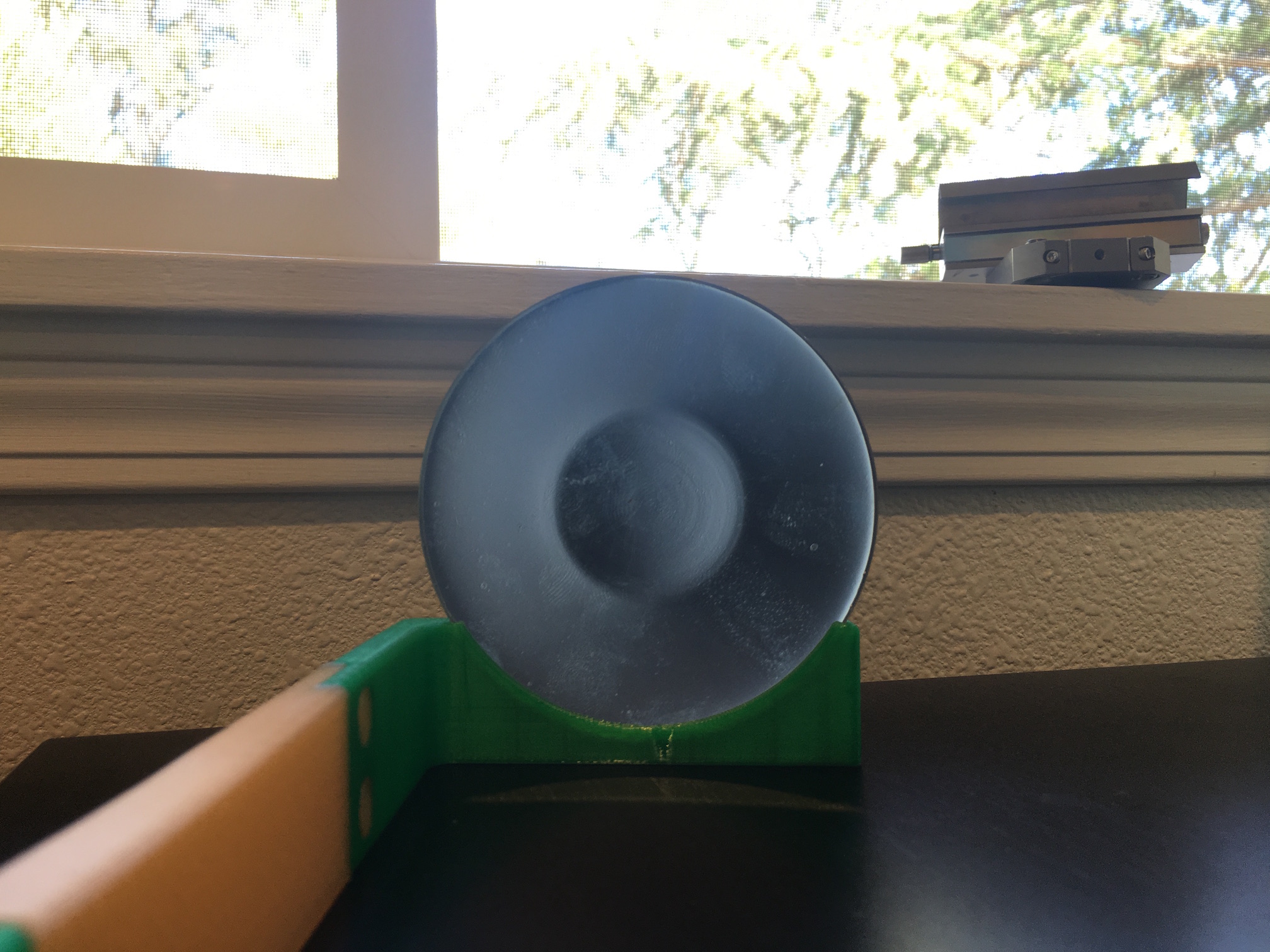}
	\includegraphics[width=50mm]{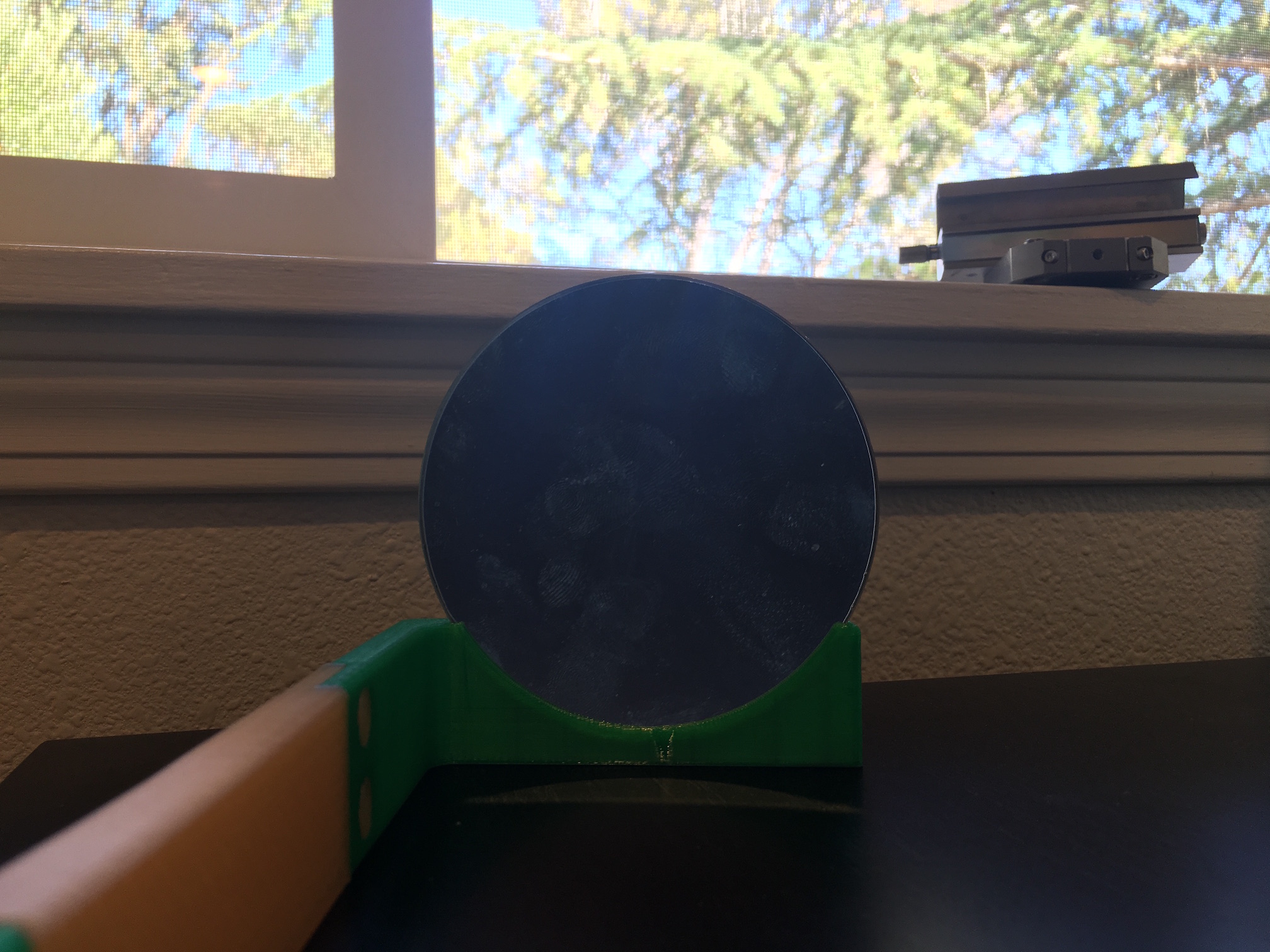}
	\caption{\label{fig:alignment_photo}Image of misaligned business card, from left to right: first, white part of card in imaged focus; second, properly aligned system; third, black part of business card in focus.}
\end{figure}

To set up and align the system, first place the mirror and phone into their respective mounts.  Then, snap the three primary members together.  Next, slide the `business card holder' into the `phone holder' member.  Then, find a business card or piece of paper.  On this, draw a vertical stripe about 1 cm wide with a marker.  The marker does not necessarily need to be black, but a black marker will result in a gray image with contrast showing up in black or white.  The business card then must be propped up against the 'business card holder' and wedged into the slot in which the business card holder slides.  The business card can then be slid laterally in order to position the contrast edge (i.e., the boundary between the stripe and the white of the business card) at the focus of the mirror imaged by the camera.  When the system is aligned, the image of the mirror on the camera will appear gray.  As you slide the business card back and forth through its correct position, the image of the mirror on the camera will transition from white to black.  When the edge between white and black is at the focus imaged by the camera, the mirror will look gray and the spherical aberration in the mirror will be apparent.  Figure \ref{fig:alignment_photo}  illustrates the alignment process.

A horizontal cutoff course image can also be used, but the height of the stripe must be adjusted by folding the card or drawing new stripes.  The single, lateral degree of freedom in the `business card holder' makes it easy to adjust a vertically striped cutoff source image, not a horizontally striped cutoff source image.   

The whole system is sensitive to vibration, which can be observed when acquiring a time series of images; the flickering results from the mirror moving relative to the segmented background.  To minimize the flickering, the system can be placed on a table with the mirror hanging over the edge of the table, and the schlieren object of interest (e.g., a lighter) can be held between the mirror and the edge of the table. 

\begin{figure}
	\includegraphics[width=50mm]{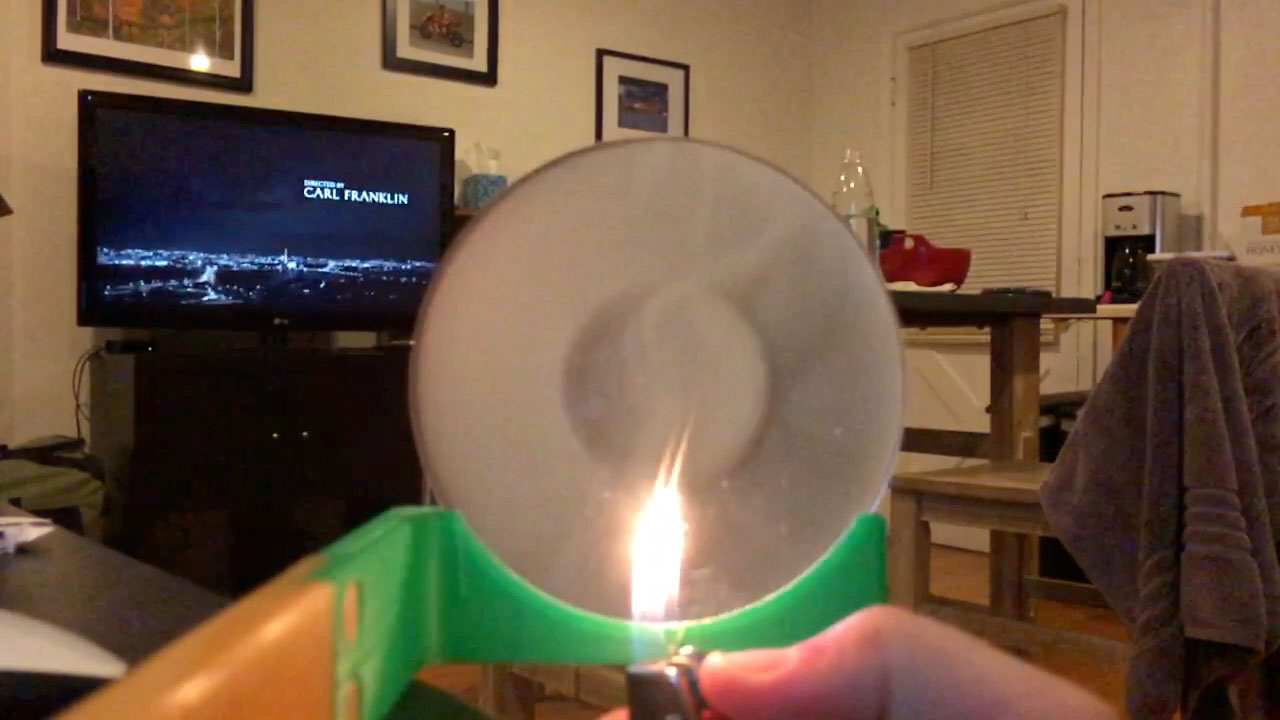}
	\includegraphics[width=50mm]{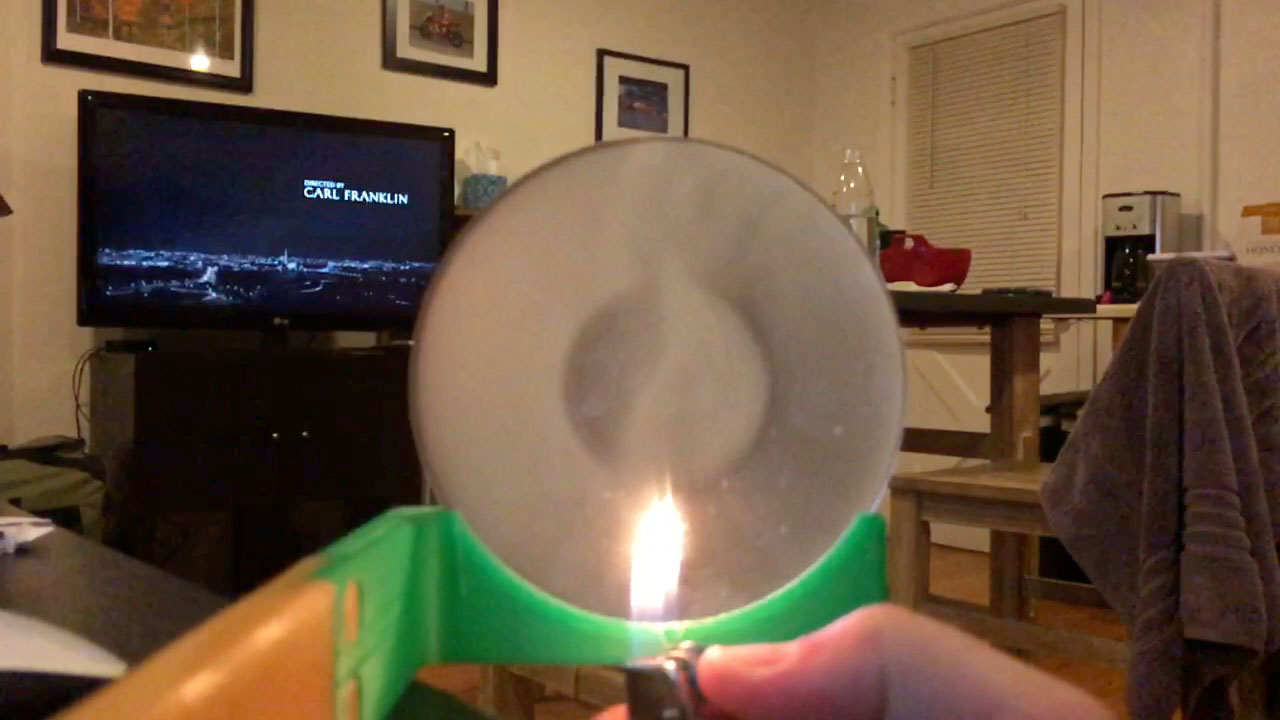}
	\includegraphics[width=50mm]{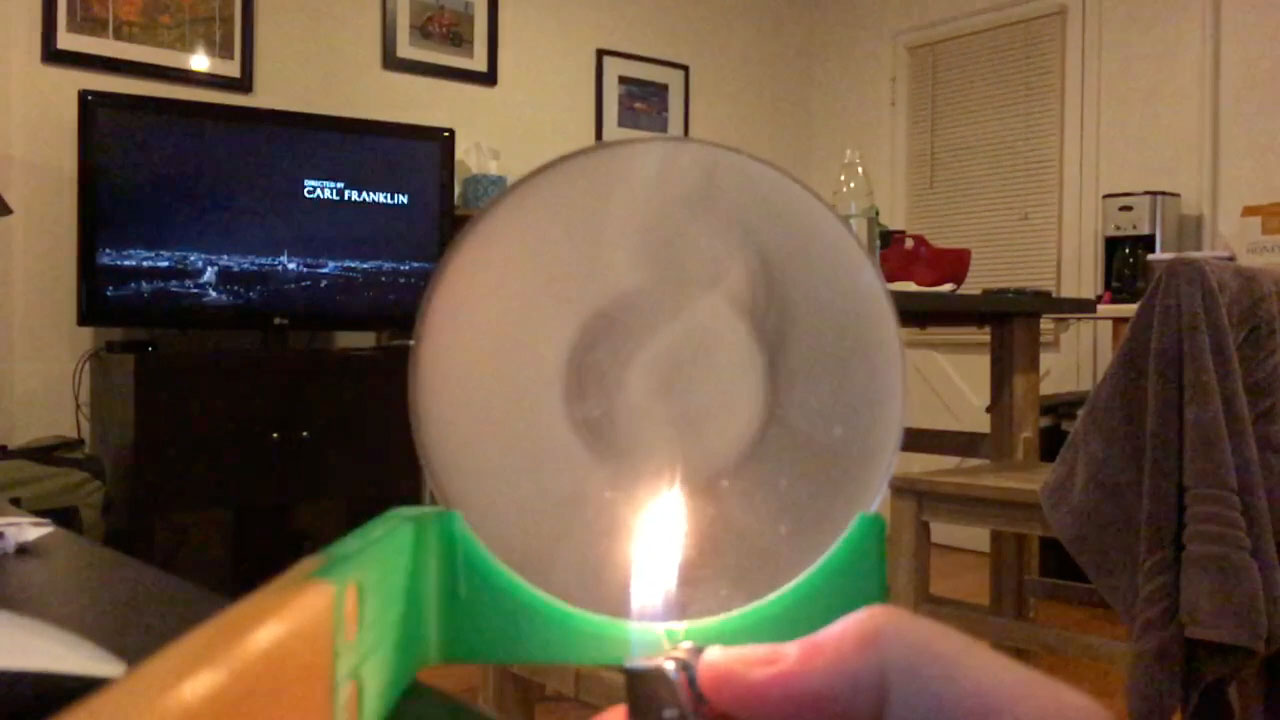}
	\caption{\label{fig:lighter}Image sequence of burning butane lighter.  The magic of OpenSchlieren does not translate well when displaying still images from the video sequence.  Video sequence is available on YouTube at \url{https://youtu.be/NQ5LPR5koTA}.}
\end{figure}

Figure \ref{fig:lighter} shows a butane lighter being lit using an iPhone 6S for the camera, which acquired images at 240 Hz, the fastest capture rate on the iPhone 6S.  This sequence is dramatic, clearly showing the spark from the flint, ignition of the butane-air mixture, and the initial plumes of hot combustion products rising from the lighter.  One could develop a lesson plan around quantifying the temperature of the flame based on the velocity of the plumes rising from the gas.  Certain assumptions about the system must be made (e.g., mixing of the combustion products and the ambient air), but this type of exercise could provide students a great opportunity to own a quantitative imaging problem and understand the uncertainties involved in making a measurement.  

\begin{figure}
	\includegraphics[width=50mm]{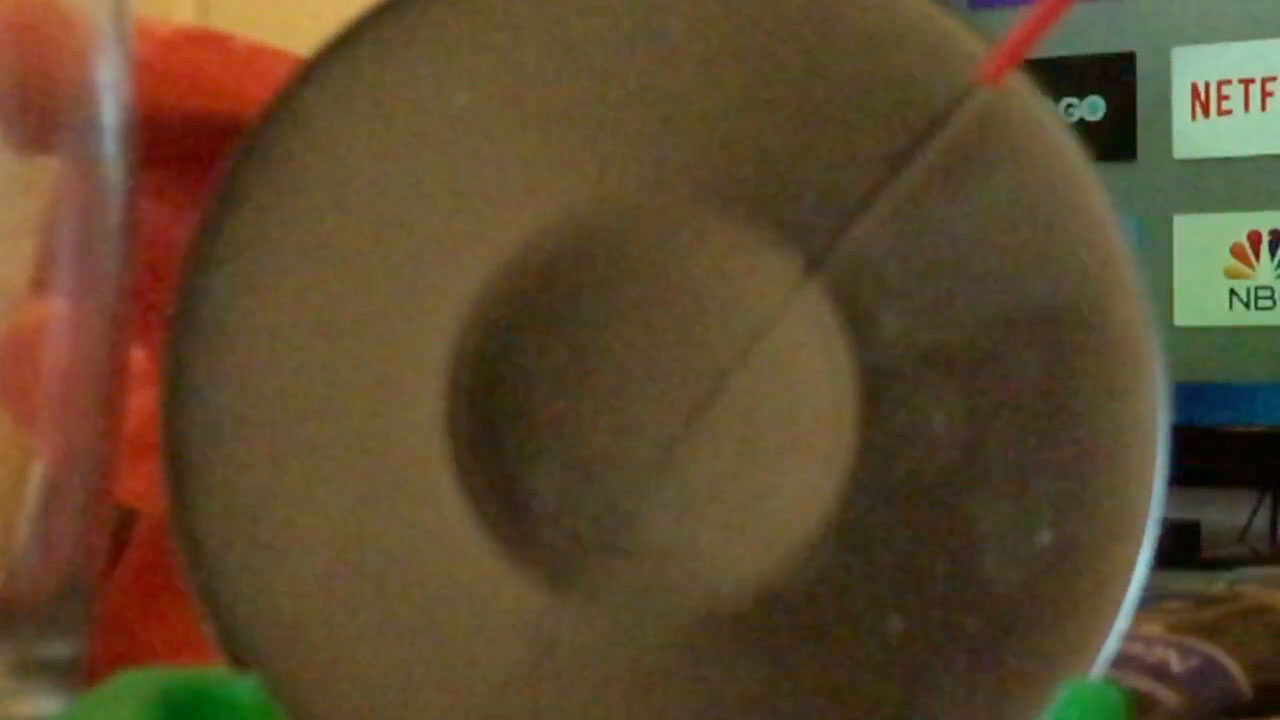}
	\includegraphics[width=50mm]{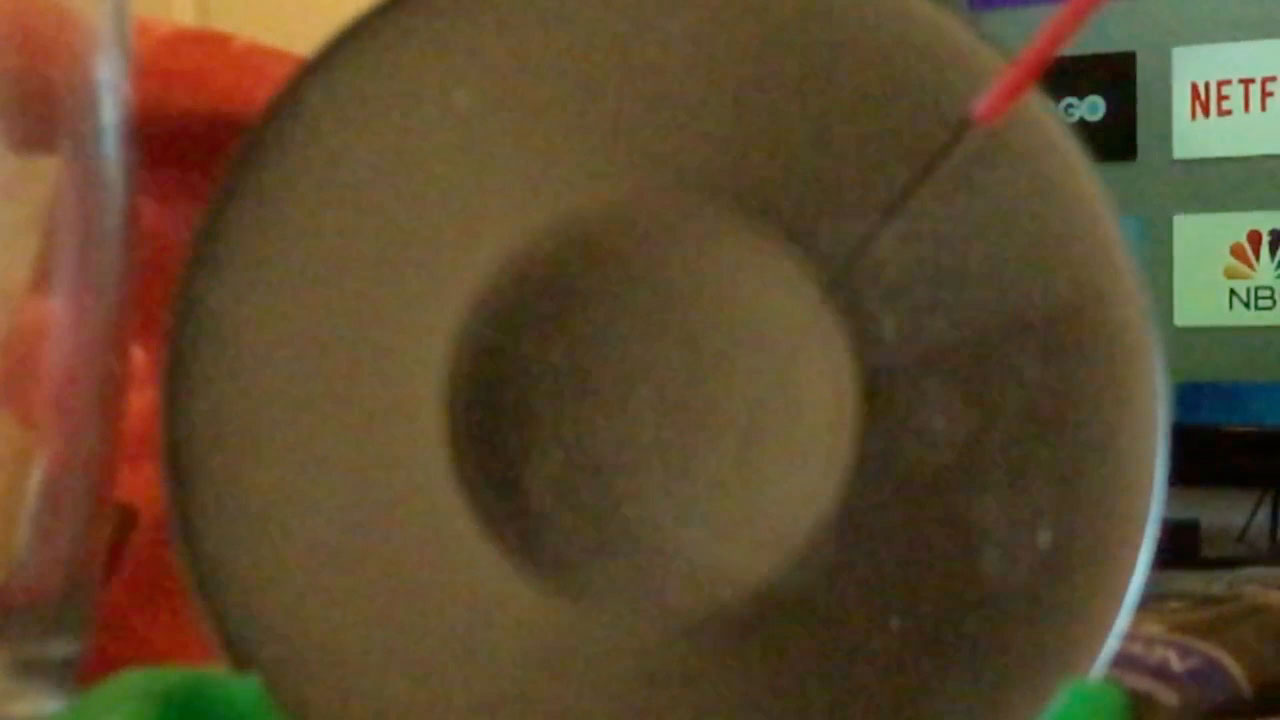}
	\includegraphics[width=50mm]{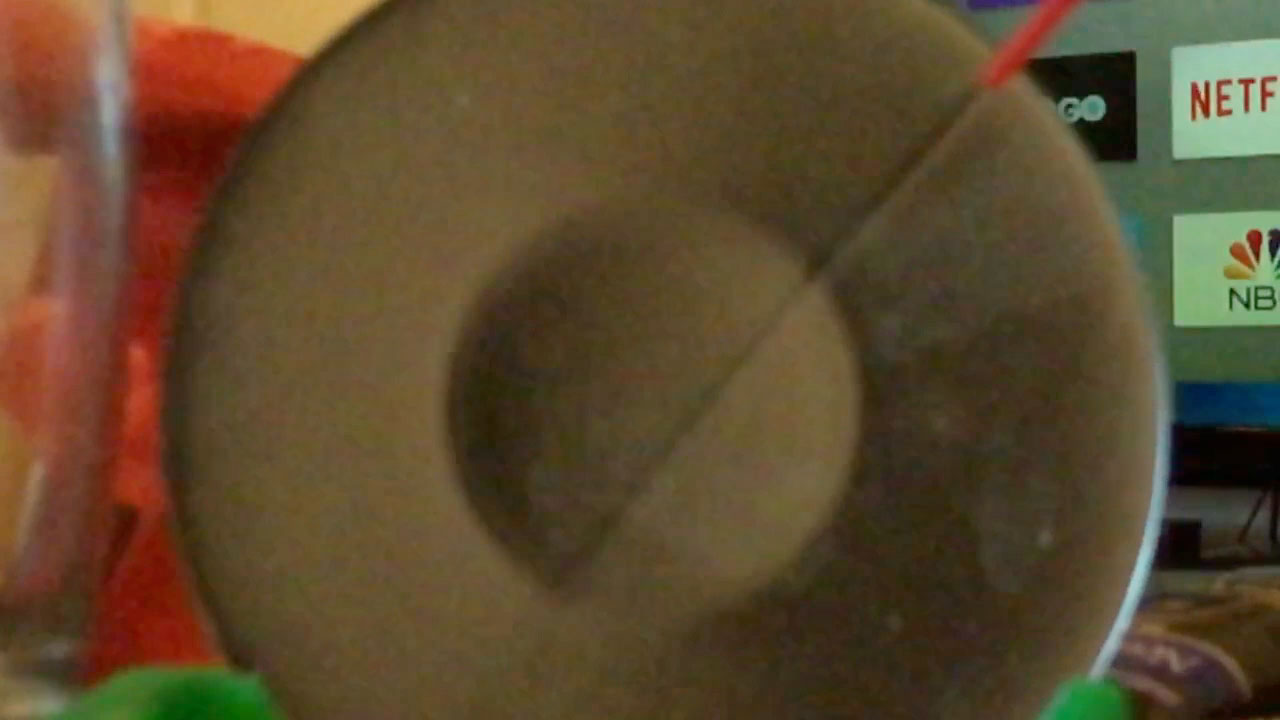}
	\caption{\label{fig:jet}Image sequence of jet exhausting from a canned air duster, highlighting the trajectory of the jet.  For this sequence, the camera was digitally zoomed to fill the field of view with the mirror.  Full video sequence is available on YouTube at \url{https://youtu.be/6DxwEAE_f9M}}
\end{figure}

Next, Figure \ref{fig:jet} shows a set of images captured of a jet of air exhausting from an air duster.  Air dusters are typically great schlieren imaging subjects, especially when working with ultra-high framing rate cameras or ultra-short exposure cameras (e.g., intensified cameras), because the turbulence in the jet can be visualized.  Here, we observe that the exposure time is too long to resolve the turbulence, but the air duster still serves a good example of illustrating the gradient direction that is observable.   In our case with a vertical stripe in the background source image, a horizontal jet (vertical density gradients) generates little contrast, whereas vertical and oblique jets generate visible contrast in the image.    

\begin{figure}
	\includegraphics[width=100mm]{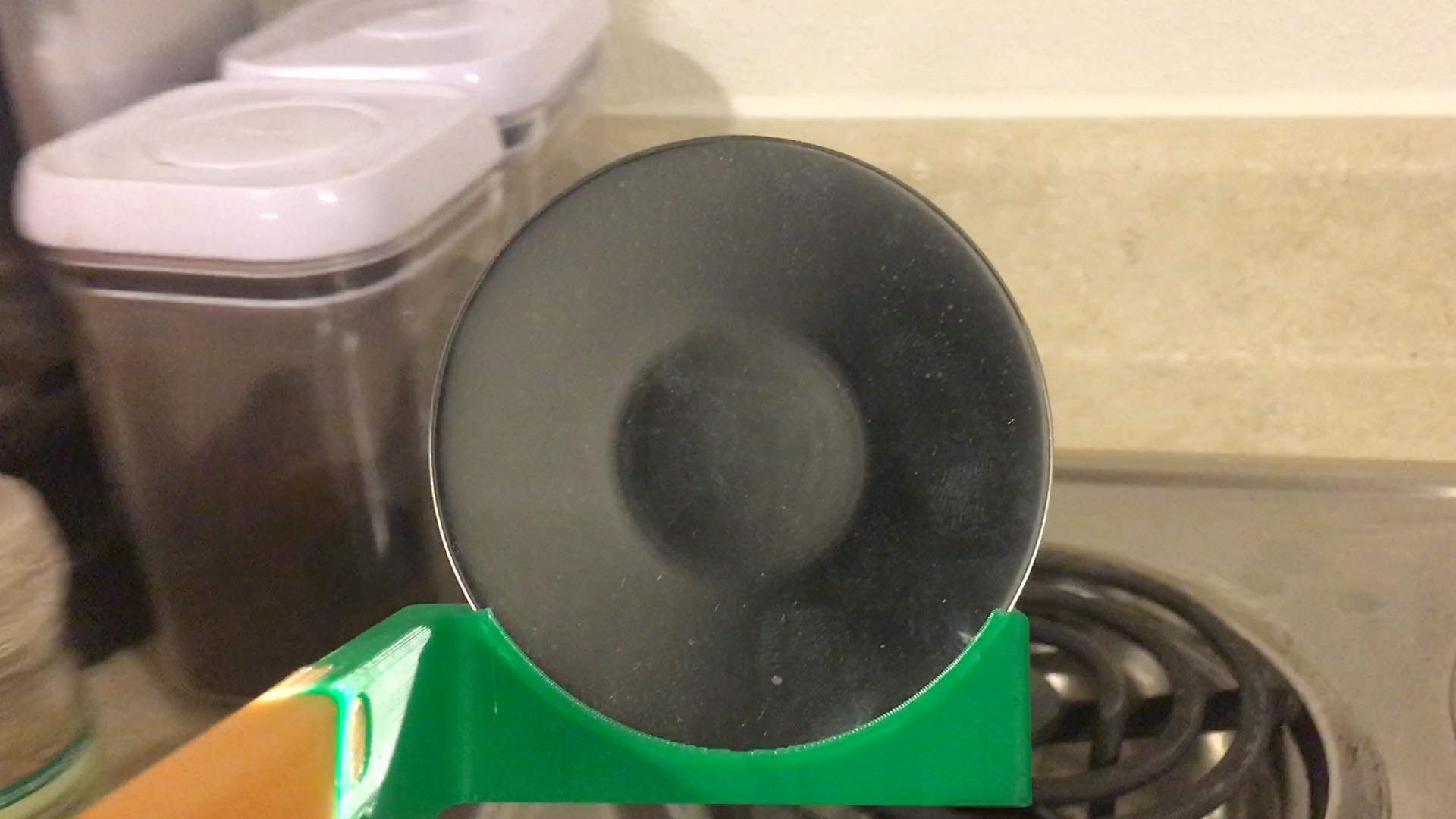}
	\caption{\label{fig:stove}A single image from a sequence acquired just before the OpenSchlieren system is held over an electric stove top.  See the video on YouTube at \url{https://youtu.be/Xn_pbWUxT9o}.}
\end{figure}

Lastly, we use OpenSchlieren to look at thermal plumes rising from an electric stove, previewed in Figure \ref{fig:stove}.  The still images translate poorly to still images due to the low sensitivity and lack of contrast generated by the system; view the video on YouTube at \url{https://youtu.be/Xn_pbWUxT9o} to see the thermal plumes.  The lack of contrast generated by the system is an opportunity for students to explore different ways to post-process images to improve contrast.  Lessons can be built around adjusting colormap limits and colors, applying unsharp mask filters, and a variety of other techniques to increase contrast and extract features from images.  Furthermore, this example highlights the susceptibility of the system to vibration.  This may be able to be improved by using a graduated cutoff (at the expensive of sensitivity) or a modified design of the system.  

\section{Conclusion}
We have presented an accessible, smartphone-based, 3D printed schlieren system.  Design considerations for the system were summarized, and we presented the first version of the system.  We used the the system to image a lighter, a canned air duster, and an electric stove, and we presented images and videos of the results.  The aim of this paper is to introduce a cheap, accessible schlieren system that uses a smartphone, and we look forward to continuing to work on this system, iterating on the design and developing curricula around a variety of topics, including basic physics, optics, rays, and imaging.  

\bibliography{openschlieren_bib.bib}
\end{document}